# Large critical current density improvement in Bi-2212 wires through groove-rolling process

**A Malagoli[1], C Bernini[1], V Braccini[1], G Romano[1], M Putti[1,2], X. Chaud[3], F Debray[3]**

[1]CNR-SPIN Corso Perrone 24, 16152 Genova, Italy
[2]Physics Department, University of Genova, via Dodecaneso 33, 16146 Genova, Italy
[3]CNRS LNCMI UJF-UPS-INSA 25, rue des Martyrs BP.166 38042 Grenoble Cedex 9, France

E-mail: andrea.malagoli@spin.cnr.it

**Abstract** Recently the interest about Bi-2212 round wire superconductor for high magnetic field use has been enhancing despite the fact that an increase of the critical current is still needed to boost its successful use in such applications. Recent studies have demonstrated that the main obstacle to current flow, especially in long wires, is the residual porosity inside these Powder-In-Tube processed conductors which develops in bubbles-agglomeration when the Bi-2212 melts.
Through this work we tried to overcome this issue acting on the wire densification by changing the deformation process. Here we show the effects of groove-rolling versus drawing process on the critical current density $J_C$ and on the microstructure. In particular, groove-rolled multifilamentary wires show a $J_C$ increased by a factor of about 3 with respect to drawn wires prepared with the same Bi-2212 powder and architecture. We think that this approach in the deformation process is able to produce the required improvements both because the superconducting properties are enhanced and because it makes the fabrication process faster and cheaper.

## 1. Introduction

Over the last years the interest about $Bi_2Sr_2CaCu_2O_x$ (Bi-2212) superconductor has been renewed thanks to its critical current density $J_C$ exceeding $10^5$ A/cm$^2$ at 4.2 K, its irreversibility field over 100 T at the same temperature [1-3] and, maybe the most important, to its peculiar characteristic to be processed as round wire, unique within the family of the cuprate superconductors. It has been already demonstrated that such properties make Bi-2212 a very competitive conductor for high field applications (> 20 T) such as magnets for nuclear magnetic resonance or accelerators [4-9], however, in this context, an enhancement of $J_C$ of two or three times seems to be still needed to boost the successful use of this superconductor in such applications.
This aim has driven the work of several groups, especially in USA – where, besides the research labs, Oxford Superconducting Technology (OST) is a company leader on Bi-2212 round wire fabrication - and in Japan and Korea. Most works were focused on studying the relationship between the fabrication process and the superconducting properties [10-15], leading to the conclusion that $J_C$ is limited by the connectivity of the filament pack thus making even more important the understanding of what controls the connectivity. More recently it has been shown that the main obstacle to current flow through the 2212 filament seems to be the presence of bubbles [12, 17, 18]. In the Powder-In-Tube (PIT) technique, some porosity remains in the powders after the Ag sheath filling [19], to avoid undesirable cracks during the cold deformation: such porosities agglomerate in bubbles as large as the filament itself in the partial-melt process, when the Bi-2212 is in the liquid phase. Such bubbles do not disappear when the solid phase forms in the fully reacted wire [18]: moreover, the pressure caused by them provokes a strong $J_C$ reduction over lengths longer than 10-15 cm [20]. The use of a 2 GPa cold isostatic pressure has successfully increased the density of the filaments of a short commercial OST wire after drawing [21], leading to a strong reduction of bubble density and size and to a doubling in the critical current value, proving that it is possible to enhance the critical current density by avoiding bubbles formation.

Strong of our experience on the Powder-In-Tube technique applied to the $MgB_2$ conductors, we developed a mechanical process based on groove-rolling able to densify the Bi-2212 phase and enhance the critical current density of the wires, which can be at the same time industrially scalable over long lengths.

Here we present a comparison between the transport properties and the microstructure of two multifilamentary Bi-2212 wires, realized by drawing and groove rolling. We found that the groove-rolled wire shows a $J_C$ increased by a factor of about 3 with respect to the drawn wire prepared with the same Bi-2212 powder and architecture. Microstructure analysis shows a higher compaction of the powder and a better uniformity of the filaments in the groove rolled sample. We therefore propose groove rolling - which is an already well-tested industrial process able to reduce the costs in terms of production time - as the method to realize Bi-2212 wires with improved superconducting properties on industrial scale.

## 2. Experimental

Two different Bi-2212 wires were realized and analysed in this work. The Powder-In-Tube method was used for both samples starting from filling an Ag tube with Nexans granulate powder. The tube outer (OD) and inner (ID) diameters were 15 and 11 mm respectively. After drawing, the obtained monofilamentary wire was hexagonal shaped and cut in 55 pieces and restacked in a second 15/11 (OD/ID) mm Ag tube. At this point the deformation process was differed for the two wires. For the sample AMBI004 the restacked tube was drawn, hexagonal shaped and cut in 7 pieces in order to be restacked again in a 10/7.5 mm (OD/ID) Ag/Mg alloy tube. After a further drawing process, a 55 x 7 filaments round wire with a diameter of 0.8 mm was obtained. For the AMBI005 sample the restacked tube with 55 filaments was worked with a groove-rolling machine. This process consists in a multiple subsequent rolling of the wire using the rollers shown in figure 1. The grooves shape is approximately square and each step produces a wire cross section reduction of about 10%. As in the drawing process, the wire was annealed several times: in this case the aim was not to avoid wire sausaging like in the drawing process, but possible cracks along the Silver sheath.

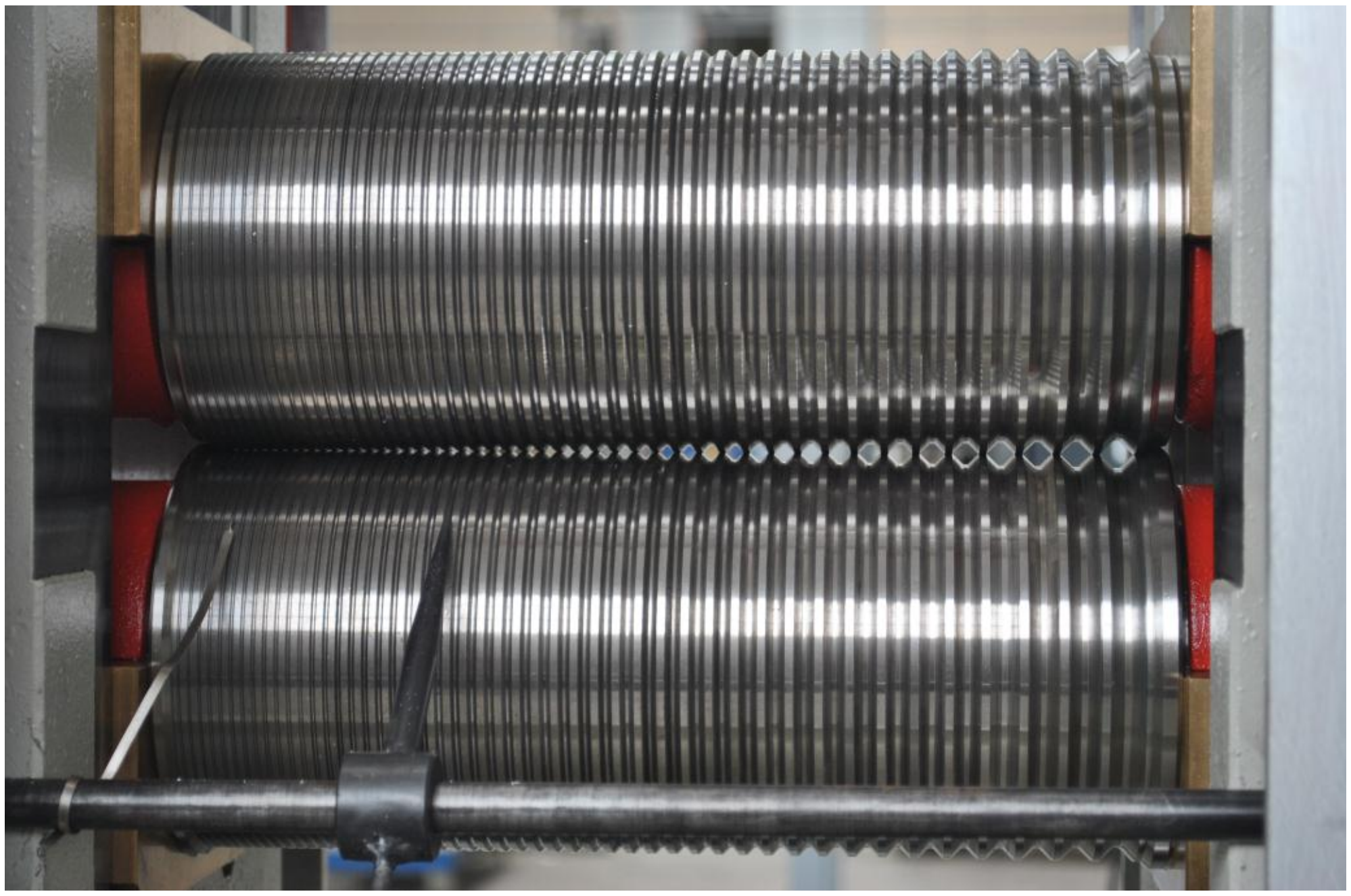

**Figure 1:** Rollers of the groove-rolling machine used for the preparation of the sample AMBI005

The so obtained wire was then hexagonal shaped, cut in 7 pieces and restacked in a 10/7.5 mm (OD/ID) Ag/Mg alloy tube. This new composite tube was groove-rolled again and a 55 x 7 filaments wire was obtained with an approximately square shape with a side of 1.1 mm (this is the present size limit of our rolling machine). Figure 2 shows images of transverse cross sections of both samples. The filling factor (f.f.) of both samples lies around 15%. To be noticed how the tubes diameter and thickness are not yet optimized.

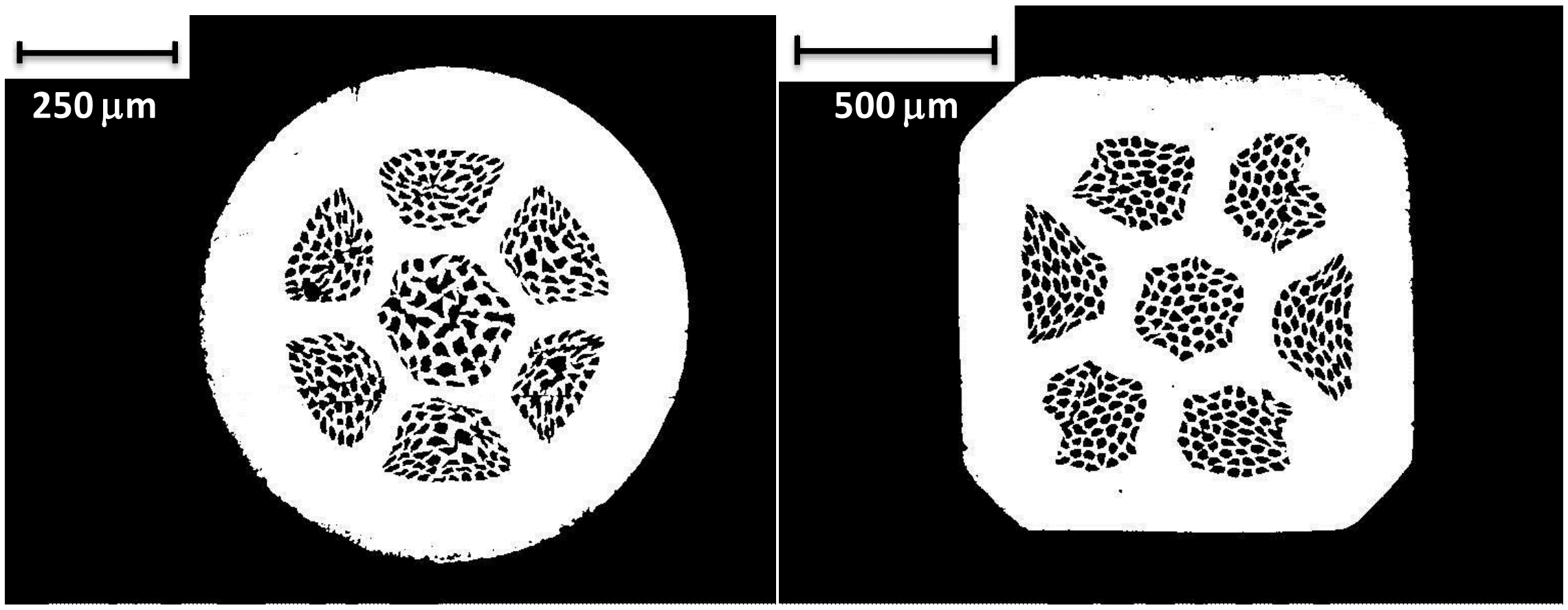

**Figure 2:** Transverse cross sections of the drawn (AMBI004 - left) and groove-rolled (AMBI005 - right) 55x7 wires

Two 15 cm long AMBI004 and AMBI005 samples were heat treated together in 0.1 MPa flowing $O_2$ in a tubular furnace with a homogeneity zone (± 0.5 °C) of 30 cm using the standard heat treatment schedule [22] shown in Figure 3.

Transverse cross sections were carefully polished with SiC paper and as a last step with a diamond film. Longitudinal cross sections were prepared through removing the Silver matrix by deep chemical etching using a solution based on ammonium hydroxide (50 vol% aqueous solution) and hydrogen peroxide (30 wt% aqueous solution), mixed in equal amounts and partly diluted by distilled water [23] in order to expose the Bi-2212 filaments. Microstructures were analysed by Scanning Electron Microscopy (SEM).

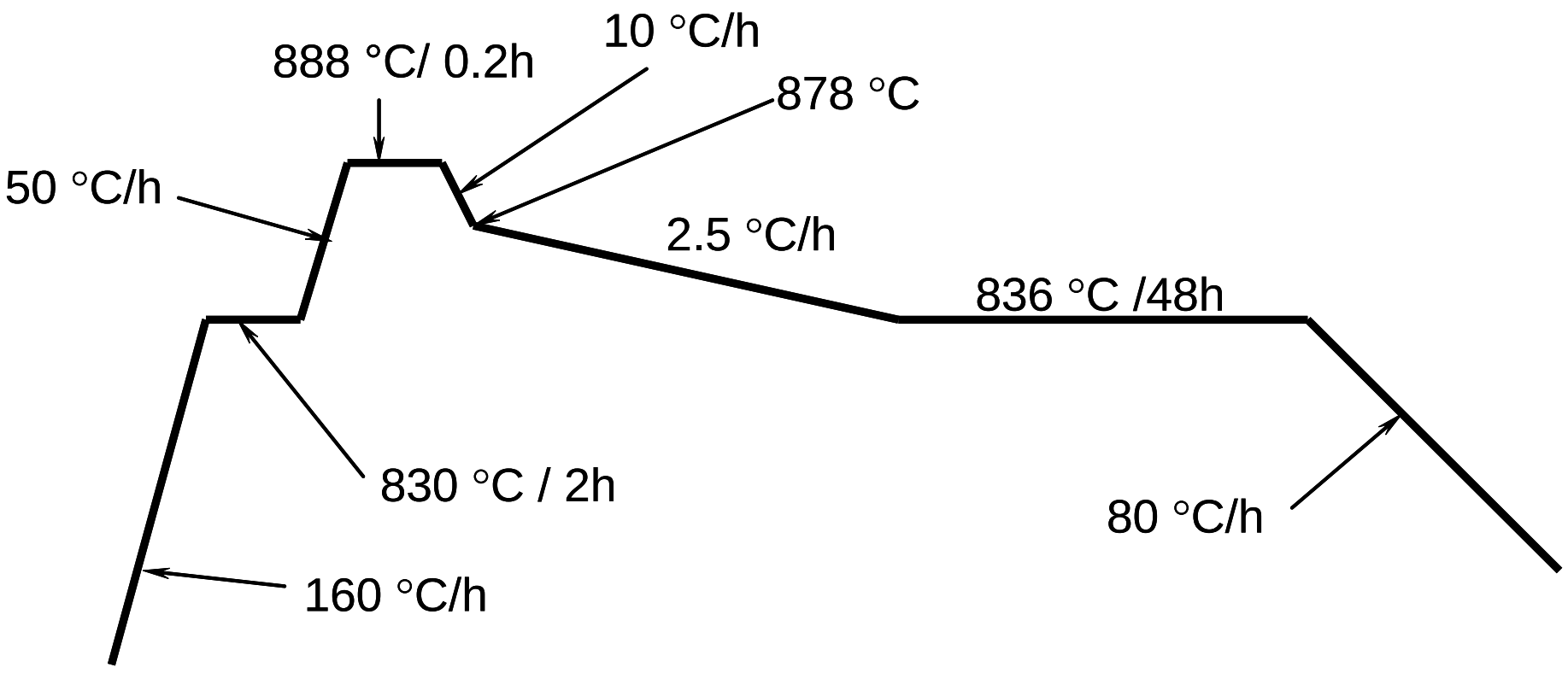

**Figure 3:** A sketch of the heat treatment performed on both the drawn and the groove-rolled wire

In order to evaluate the mass density of the two wires, the transverse surfaces at the ends of each sample were carefully polished until they matched a flat reference surface. Then their length was measured with a micrometer with an accuracy of 0.001 mm while their mass was measured with a balance with an accuracy of 0.01 mg. Their cross-sections were imaged in an optical microscope which gave high contrast digital images with a resolution of ~1 µm over the whole wire cross-section. These images were analysed with ImageJ so as to extract the area fraction of the Ag matrix and filament packs for evaluation of the effective area of the filaments.

To estimate the starting powder granulometry, several images of the 2212 powders were taken by SEM and afterwards analysed by software measuring the grains size. For this analysis the powder was uniaxially pressed in a few millimetre thick pellet. This allowed us to observe the powder grains on the same plane and thus better distinguish and measure their size. We performed such measurement on about 2000 different grains.

Transport critical currents were measured on 5 cm long samples by means of a four-probe system using a 7 T split-coil magnet at 4.2 K whose field was applied perpendicular to the wire axis. A criterion of 1 µV/cm was used. The critical current density $J_C$ was calculated taking into account the oxide filaments area measured by image analysis on the unreacted wire as it is used in literature [12].

## 3. Results

Figure 4 shows $J_C$ at 4.2 K for the two samples AMBI004 and AMBI005.

As well shown in the inset - where the measured V-I curves at 7 T are reported - the critical current $I_C$ of AMBI005 exceeded by a factor 8 those of AMBI004. $J_C$ values evaluated at 6 and 7 T taking into account the different superconducting cross-section of the wires show a remarkable improvement of more than three times by changing the deformation process.

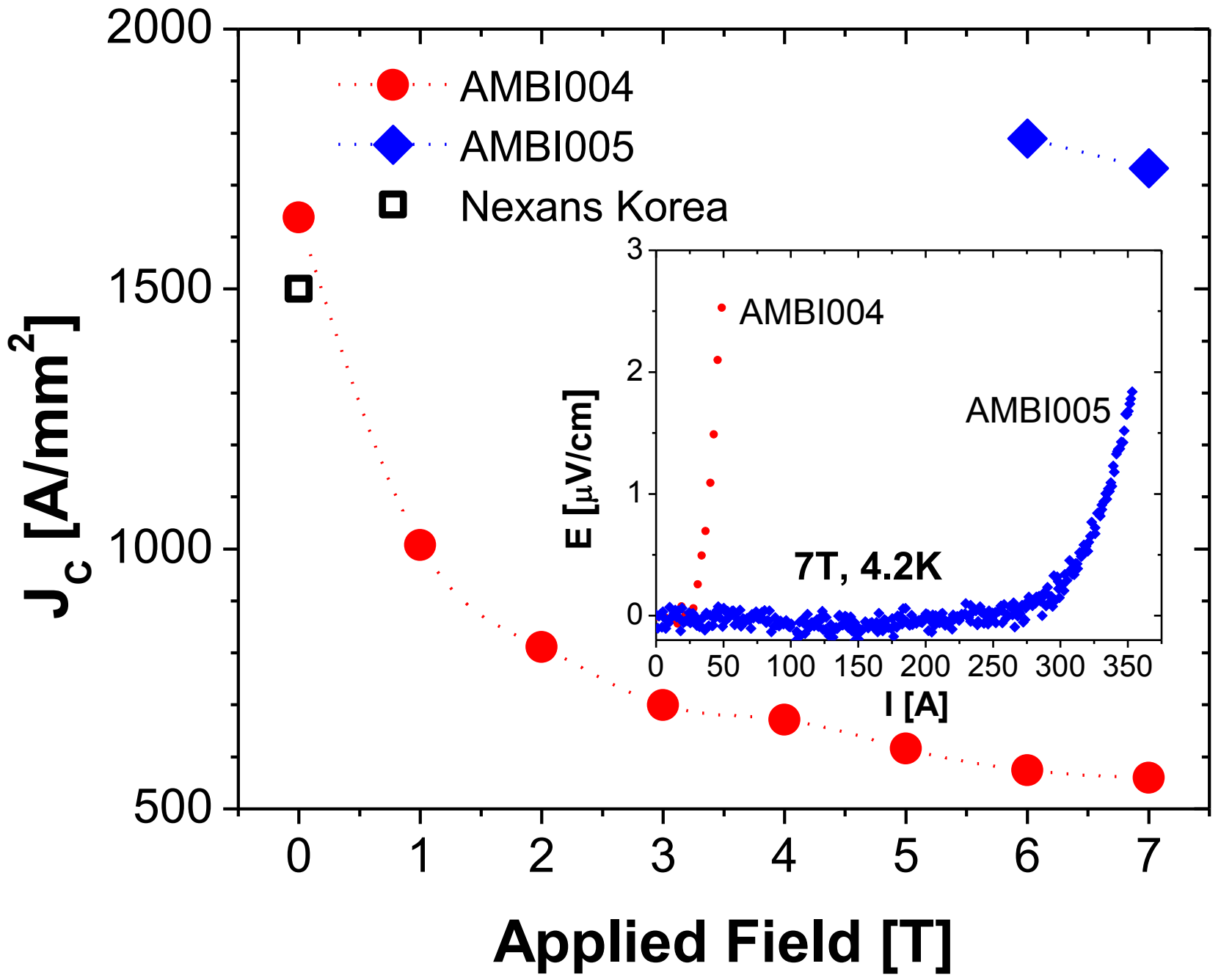

**Figure 4:** $J_C$ vs. B at 4.2 K for the drawn (AMBI004) and groove-rolled sample (AMBI005) and comparison with the value in self-field of a similar drawn 55x7 wire from ref. [16]. In the inset, the V-I curves are shown for the two samples at 7 T.

These results are compared (in self-field) with the value of a 55 x 7 wire fabricated by Nexans Korea reported in [16]. This is an optimized wire and the most similar to AMBI004, concerning powders and especially filament architecture, found in the literature: therefore such comparison is particularly useful to test the reliability of our drawn wire. The values in self-field for the two drawn wires are very similar, even if the filling factor was not optimized in AMBI004, being about 15% instead of the 20% reported in [16] for the same filaments design.
To better understand in which way the differences in the deformation process act on the superconducting properties we thoroughly analysed the granulometry of the powders and the structure of the wires before and after the reaction. In figure 5 the evaluated grain size distribution of the Nexans Granulate is shown (left) together with an example of how such powders look at SEM (right). The peak corresponds to a grain size value of about 0.65 μm but what is particularly interesting is the presence of large grains or agglomerates of them which are of 2-4 μm in size, quite evident and anything but not rare as shown by the SEM image and the several and not negligible peaks in the distribution.

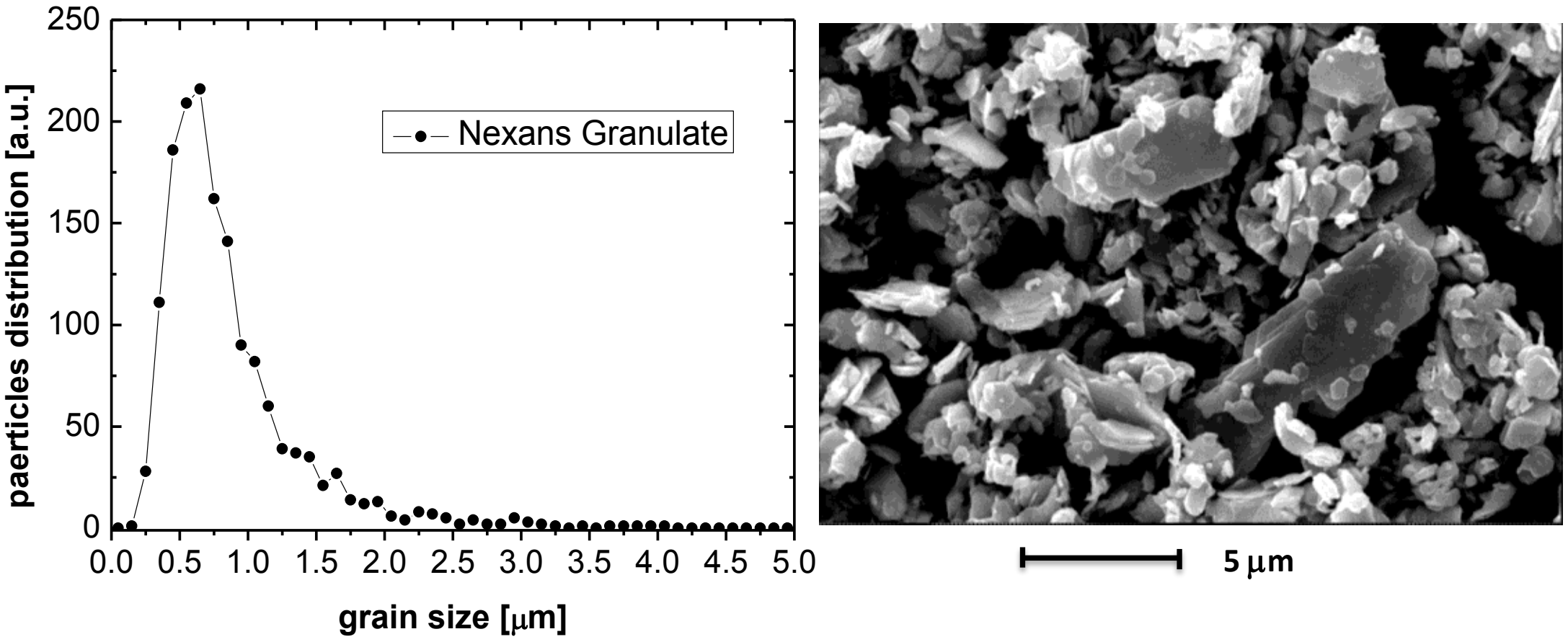

**Figure 5:** Grain size distribution (left) of the Nexans granulate used to fabricate the wires extracted from SEM analysis (right)

Figure 6 shows SEM images of the longitudinal cross section of the unreacted AMBI004 (on the left) and AMBI005 (on the right) samples as drawn and as groove rolled respectively. The images of both wires were taken at two different magnifications. Comparing the images we can see that the filaments, although different in size, in AMBI005 are more regular and uniform with respect to those in AMBI004. Furthermore, despite AMBI004 underwent a higher reduction ratio, large grains agglomerates are still quite evident in its filaments (marked by a circle in the picture) which are even irregularly deformed, while no evidence of them seems to appear in AMBI005. Finally, looking at the higher magnification images, in AMBI005 the filaments look more dense and compacted.
The images of the longitudinal cross sections of the reacted AMBI004 (on the left) and AMBI005 (on the right) samples are shown in figure 7. From the images comparison at several grades of magnification, a higher compaction and uniformity can be observed in AMBI005 filaments than in AMBI004 ones. The large filaments merging, which occurs in both samples and makes the filaments difficult to be distinguished, is typical of these Bi-2212 wires and has already been well described elsewhere [12], but what is more important here is that in AMBI004 more void spaces can be observed within the 2212 filaments and the filaments themselves look more disordered then in AMBI005. Being for the two samples the initial powder density and the Ag sheaths thickness the same, we can gather that the rolling process has been able to densify more the 2212 powder with respect to what the drawing process can do.

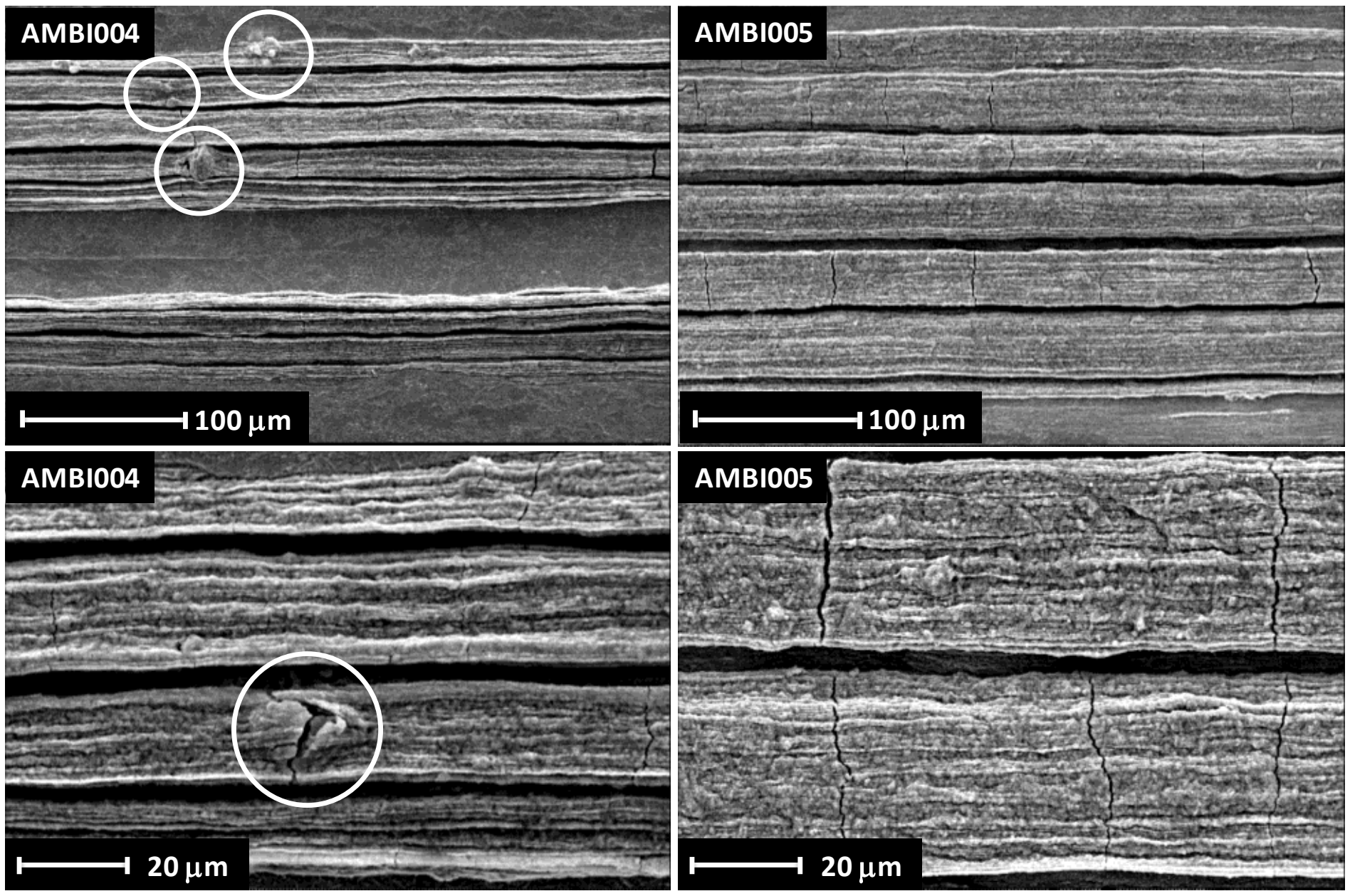

**Figure 6:** SEM images of the longitudinal cross section of the unreacted AMBI004 (left panels) and AMBI005 (right panels) samples as drawn and as groove rolled respectively. Images at two different magnifications are shown. The white circles mark the grains agglomerates not crushed by cold deformation.

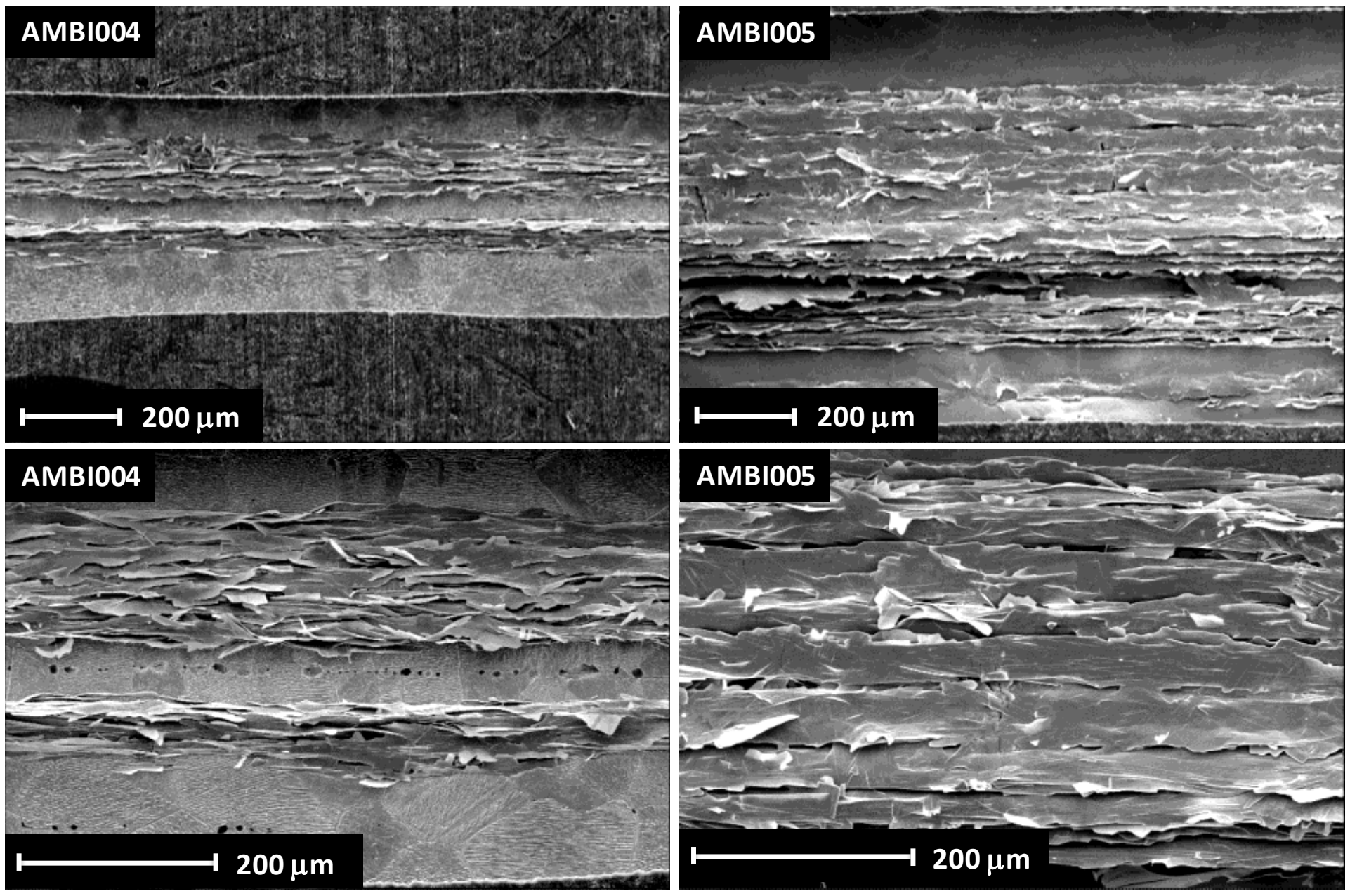

**Figure 7:** SEM images of the longitudinal cross section of the reacted AMBI004 (left panels) and AMBI005 (right panels) wires after the melt. Images at two different magnifications are shown.

## 4. Discussion

Recent studies [17, 18, 20, 21] have shown already how the bottleneck to get the desired increase in $J_C$ in the Bi-2212 wires to make them definitely appealing for applications in high magnetic field is the presence of bubbles in the final wire, which is related to the density of the superconducting phase inside the wire at the end of the deformation process. Of course one cannot increase the initial powder density at discretion, being limited by the possibility of deforming the wire [19]. The solution therefore lies in succeeding in fabricating an as-deformed conductor with the highest possible density. The idea of exploiting the very high power of the groove-rolling process in terms of powder compaction came from our previous work on $MgB_2$ wires [24].

We analysed two wires fabricated through rolling and drawing having a quite simple architecture – especially if compared with the up-to-date industrial multifilamentary samples – and in particular not optimized in terms of sheath thickness and fill factor. However we got an enhancement in $J_C$ of more than a factor 3 through rolling with respect to the drawn wire. Being our drawn wire very similar both in the architecture and in terms of performance (see the critical current comparison in figure 4) to a wire produced by Nexans Korea, we are confident that our drawn wire shows a reasonable $J_C$ value, and therefore we believe that the comparison is reliable and the enhancement obtained through rolling is real.

It is known that the best performing industrial wires are those optimized and manufactured by OST: they present higher $J_C$ values at the same temperature and magnetic field than our drawn wire. However, it is known as well that the comparison between Bi-2212 wires fabricated through different processes, and with different architectures and filament size is not so simple and straightforward. For example, the 85 x 7 filaments OST wire has a much higher $J_C$ with respect to our AMBI004, while for the 27 x 7 OST wire $J_C$ is lower [25]. Similarly being the filaments in AMBI004 smaller than in AMBI005, the comparison could appear not much valuable, but the aim of the present work is just to compare the effects led by two different techniques to exactly the same samples in terms of powder, initial sheaths thickness etc. Therefore we think that this comparison is still valuable and significant even because groove-rolling results not only in an enhancement of $J_C$ but also in a higher density of the Bi-2212 filaments and in their more uniformity.

The higher compaction and better uniformity of the filaments obtained through rolling has been confirmed by the SEM analysis reported in the previous section. Already from the images of the unreacted wires we clearly see denser and more uniform filaments in the groove-rolled sample (figure 6): we believe that such deformation process is more effective in crushing down the powders, even the larger agglomerates observed in figure 5. Having a higher compacted powder before the partial-melt process produces filaments with less voids and a higher uniformity also in the full reacted wire, as shown in figure 7.

In order to give a further support to this result, we evaluated the Bi-2212 mass density in the two full-reacted wires as described in Section 2. The Bi-2212 mass density ($\rho_{Bi}$) calculation is based on the image analysis of the sample cross sections and the theoretical value of the Ag density ($\rho_{Ag}$). The formula used for such calculation is the following:

$$\rho_{Bi} = \frac{m_{tot} - \rho_{Ag} * Vol_{Ag}}{Vol_{filaments}}$$

where $m_{tot}$ is the measured total mass of the sample, $Vol_{Ag}$ and $Vol_{filaments}$ are respectively Ag volume and the volume composed by Bi-2212 + void space obtained by the measured sample length and the imaged cross sections. We found $\rho_{Bi}$ = 3.1 g/cm$^3$ and $\rho_{Bi}$ = 4.1 g/cm$^3$ for the samples AMBI004 and AMBI005 respectively: the groove-rolled sample shows a higher density, as expected from SEM analysis. However, the density calculation could be affected by uncertainty about the value of the Ag mass density after heat treatment and is related to the image analysis accuracy: it is therefore difficult to guarantee that the absolute values are correct, but we think that the comparison between them can give a reliable indication to support our conclusions.

From an industrial point of view the rolling process is particularly appealing: besides the strong increase in $J_C$, it is less time-consuming – in fact, for example, it does not need the swaging steps as in drawing process - and most important it does not present the risk of filament sausaging. Therefore, the annealing steps can be reduced: they are only needed to avoid cracks in the sheath due to such a heavy deformation process, but being the Ag very ductile the few annealing steps we performed were enough to avoid any crack. All these aspects lead to a global costs reduction.
Furthermore, its scaling up over wires lengths of the order of the km is really straightforward: this kind of process is used for example by Columbus Superconductors to produce $MgB_2$ multifilamentary wires in lengths > km [26].

## 5. Conclusions

With the aim of fabricating Bi-2212 round wires with a higher powder compaction before the melt in order to diminish the bubble density, we fabricated two wires with the same 55 x 7 architecture through drawing and groove-rolling, and compared their microstructure before and after the melt and their performances in terms of critical current. We found that the only different deformation leads to a three-fold improvement in the $J_C$ values at 4.2 K, 7 T. The reason lies in the higher compaction reached through groove rolling, as evidenced by the SEM analysis and mass density evaluation.
We have shown that the groove-rolling procedure can be successfully applied to the fabrication of Bi-2212 wires. Still there is a large margin of improvement and refinement: the next steps will be the optimization of the architecture – increase the number of filaments and fill factor - and of the cold deformation steps, for example in terms of single cross-section reduction step and speed of the process. The application to long length wires and the eventual industrialization of the process is quite immediate and particularly appealing being even faster and cheaper than the present fabrication procedure.

**Acknowledgements**
We thank Mark Rikel for the useful discussion.